\documentclass[aps,prd,preprint,nofootinbib,11pt]{revtex4}
\usepackage{amsfonts}
\usepackage{mathrsfs}
\usepackage{graphicx}
\usepackage{amsmath}
\usepackage{amssymb}
\usepackage{subfigure}
\usepackage{epsfig}
\usepackage{graphicx}
\usepackage{color}
\usepackage{multirow}
\newcommand{\bqa}{\begin{eqnarray}}
\newcommand{\eqa}{\end{eqnarray}}
\newcommand{\beq}{\begin{equation}}
\newcommand{\eeq}{\end{equation}}
\allowdisplaybreaks[1]
\graphicspath{{fig/}{dia/}} \DeclareGraphicsExtensions{.eps}

\begin{document}

\title{Gluon-pair-Creation Production Model of Strong Interaction Vertices\\[0.7cm]}

\author{Bing-Dong Wan$^1$\footnote{wanbingdong16@mails.ucas.ac.cn} and Cong-Feng Qiao$^{1,2}$\footnote{qiaocf@ucas.ac.cn, corresponding author}\vspace{+3pt}}

\affiliation{$^1$ School of Physics, University of Chinese Academy of Science, Yuquan Road 19A, Beijing 10049 \\
$^2$ CAS Center for Excellence in Particle Physics, Beijing 10049, China}

\author{~\\~\\}

\begin{abstract}
\vspace{0.3cm}
By studying the $\eta_c$ exclusive decay to double glueballs, we introduce a model to mimic phenomenologically the gluon-pair-vacuum interaction vertices, namely the $0^{++}$ model. Based on this model, we study glueball production in pseudoscalar quarkonium decays, explicitly $\eta_c \to f_0(1500)\eta(1405)$, $\eta_b\to f_0(1500)\eta(1405)$ and $\eta_b\to f_0(1710)\eta(1405)$ processes. Among them $f_0(1500)$ and $f_0(1710)$ are well-known scalars possessing large glue components and $\eta(1405)$ is a potential candidate for pseudoscalar glueball. The preliminary calculation results indicate that these processes are marginally accessible in the presently running experiments BES III, BELLE II, and LHCb.

\end{abstract}
 \maketitle
\newpage

\section{Introduction}

According to the theory of strong interaction, the Quantum Chromodynamics (QCD) \cite{QCD}, gluons have self-intraction, which suggests in some sense the existence of glueball. The search for glueballs has experienced a long history, however the existence evidence is still vague. Being short of reliable glueball production and decay mechanisms makes the corresponding investigation rather difficult. Another hurdle hindering the glueball searching lies in the fact that usually glueballs mix heavily with the quark states, somehow with the exception of exotic glueballs \cite{exoticglueballs}.

The scalar glueballs which have the quantum numbers $J^{PC}=0^{++}$ are suggested to be the lightest glueballs by lattice calculation and its mass is around $1600-1700$ MeV with an uncertainty of about $100$ MeV \cite{Lee:1999,Bali:1993,Morningstar:1997,Chen:2006}. Experimentally, there exist three isosinglet scalars $f_0(1370)$, $f_0(1500)$ and $f_0(1710)$ around $1600-1700$ MeV. The absence of the $\gamma\gamma\to K\bar{K}$ or $\pi^+\pi^-$ mode through $f_0(1500)$ excludes the possibility of a large $n\bar{n}$ content within $f_0(1500)$ \cite{Acciarri:2001a,Barate:2000}. On the other hand, the $f_0(1500)$ has a small $K\bar{K}$ decay branching rate \cite{Abele:1996,Abele:1998,Barberis:1999,Janowski:2014ppa}, implies its main content is hardly to be $s\bar{s}$. Various peculiar natures suggest that $f_0(1500)$ might be a scalar glueball, or glue rich object \cite{Amsler:1995a} . In a large mixing model, as discussed in Refs. \cite{Amsler:1995a,Close:2005,He:2006,Yuan:2011}, glue is shared between $f_0(1370)$, $f_0(1500)$ and $f_0(1710)$. The $f_0(1370)$ is mainly constructed of $n\bar{n}$, the $f_0(1500)$ is thought to be glue predominant, and the $f_0(1710)$ has a large content of $s\bar{s}$ ingredient.

Evidence for pseudoscalar $0^{-+}$ glueballs is still weak \cite{Crede:2009}. $E(1420)$ and $\iota(1440)$ observed by Mark II are early candidates of pseudoscalar glueballs proposed in Refs. \cite{Donoghue:1981,Chanowitz:1981,Ishikawa:1981,Lacaze:1981}. However, $E(1420)$ is later reconsidered as the $1^+$ meson and renamed $f_1(1420)$, while $\iota(1440)$ is still thought to be a pseudoscalar, now known as $\eta(1405)$ \cite{Masoni:2006}. $\eta(1405)\to\eta\pi\pi$ was observed at BES II in $J/\psi$ decay \cite{Ablikim:2011}, and was confirmed in $\bar{p}p$ annihilation \cite{Amsler:1995b}. It should be noticed that $\eta(1405)$ was observed in neither $\eta\pi\pi$ nor $K\bar{K}\pi$ channels in $\gamma\gamma$ collisions by L3 \cite{Acciarri:2001b}, which implies $\eta(1405)$ has a large glue component since glueball production is suppressed in $\gamma\gamma$ collision. It is worth mentioning that the quenched Lattice and QCD Sume Rule calculation predict that the $0^{-+}$ glueball mass might be above $2$ GeV \cite{Bali:1993,Morningstar:1999,HQZ}, though Gabadadze argued that the pseudoscalar glueball mass in full QCD could be much less than the quenched Lattice result in Yang-Mills theory \cite{Gabadadze:1997zc}. Furthermore, despite $\eta(1405)$ fitting well the fluxtube model \cite{Faddeev:2004} and roughly the $\eta$-$\eta'$-$G$ mixing calculations \cite{Cheng:2009}, a recent triangle singularity mechanism analysis tells that $\eta(1405)$ and $\eta(1475)$ might be the same state \cite{Du-and-Zhao}. About more properties of pseudoscalar glueballs, readers may refer to recent studies, e.g. \cite{Eshraim:2012jv,Brunner:2016ygk}.

\begin{figure}[htbp]
\begin{center}
\includegraphics[scale=0.46]{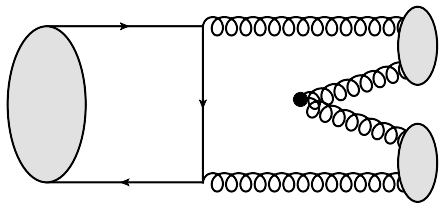}
\caption{Schematic diagram for glueball production in $\eta_c$ decay in $0^{++}$ model.}\label{decayGPC}
\end{center}
\end{figure}

In this paper, motivated by studying the glueball production and decay mechanisms, we discuss the glueballs production in $\eta_c$ decay by introducing a model for the gluon-pair-vacuum interaction vertices, namely the $0^{++}$ model, as shown in the Fig. \ref{decayGPC}. We assume the gluon pair is created homogeneously in space with equal probability. Comparing to the $^3P_0$ model \cite{Micu:1969,yaouanc:1973,yaouanc-book,Ackleh:1996,Luo:2009,Sun:2014,Burns:2014,Blundell:1996,Segovia:2012,Yang:2009fj}, which models the quark-antiquark pair creation in the vacuum, we formulate an explicit vacuum gluon-pair transition matrix and estimate the strength of the gluon-pair creation. Employing the $0^{++}$ model, we then investigate the $\eta_c$ and $\eta_b$ decays to scalar and pseudoscalar glueballs. Based on our knowledge about glueballs, we take $f_0(1710)$ and $f_0(1500)$ as scalar glueball candidates, and $\eta(1405)$ as a pseudoscalar glueball candidate. The corresponding decay widths and branching fractions are calculated.

The rest of the paper is arranged as follows. After the introduction, we construct a model for gluon-pair-vacuum interaction vertices in Sec.\ref{construction}. The partial widths of $\eta_c \to f_0(1500)\eta(1405)$, $\eta_b\to f_0(1500)\eta(1405)$ and $\eta_b\to f_0(1710)\eta(1405)$ are evaluated in Sec.\ref{decay}. Last section is remained for summary and outlooks.

\section{Construction of the $0^{++}$ model}\label{construction}

In quantum fields theory, the physical vacuum is thought of the ground state of energy, with constant particle fields fluctuations. Therefore, there are certain probabilities of quark pairs and gluon pairs with vacuum quantum numbers showing up in the vacuum. It is reasonable to conjecture that gluon pairs would be created with equal amplitude in space, just like the quark-antiquark pairs do in $^3P_0$ model. Created from the vacuum, the gluon pairs hence possess the quantum numbers $J^{PC}=0^{++}$.

We may argue the soundness of the $0^{++}$ scheme like this: in the language of Feynman diagram, the dominant contribution to the vacuum-gluon-pair coupling may stem from the processes where additional two gluons are produced from either parent meson or first two gluons. It should be noted that although by naive order counting of strong coupling one may presumably say these processes are dominant, however, in fact, the nonperturbative effect may impair this analysis. The most straightforward way to configure the vacuum-gluon-pair coupling is to attribute various contributions to an effective constant, in analogous to $^3P_0$ model. This is somewhat like the case of hadron production, where only limited hadron production processes have been proved to be factorizable, and all others are usually evaluated by assumptions or models.

In what follows we investigate glueball pair production in pseudoscalar quarkonium decay by means of the $ {0}^{++} $ model. The transition amplitude of $\eta_c$ exclusive decay to double glueballs for instance, as shown in Fig. \ref{decayGPC}, can be formulated as
\begin{eqnarray}
\langle G_1G_2|T|\eta_c\rangle = \gamma_g \langle G_1G_2 |T_2 \otimes ({G_{\rho\sigma}^c G^{c\rho\sigma}})| \eta_c\rangle\ . \label{transtiion}
\end{eqnarray}
Here, $G_1$ and $G_2$ represent glueballs; $\gamma_g$ denotes the strength of gluon pair creation in the vacuum, which in principle can be extracted by fitting to the experimental data. The ${G_{\rho\sigma}^c G^{c\rho\sigma}}$ term creates the gluon pair in the vacuum. $T_2$ is the transition operator for $\eta_c$ annihilating to two gluons. The $|\eta_c\rangle$ and $T_2$ can be expressed as
\begin{eqnarray}
|\eta_c \rangle &=& \sqrt{2 E_{\eta_c}} \int \rm d^3 \mathbf{k}_c\rm d^3 \mathbf{k}_{\bar{c}} \delta^3\left(\textbf{K}_{\eta_c}-\mathbf{k}_c-\mathbf{k}_{\bar{c}}\right)\nonumber\\
&\times& \sum_{M_{L_{\eta_c}},M_{S_{\eta_c}}} \left\langle L_{\eta_c} M_{L_{\eta_c}} S_{\eta_c} M_{S_{\eta_c}} | J_{\eta_c} M_{J_{\eta_c}} \right\rangle \psi_{n_{\eta_c} L_{\eta_c} M_{L_{\eta_c}}}\left(\mathbf{k}_c,\mathbf{k}_{\bar{c}}\right) \chi^{c \bar{c}}_{S_{\eta_c} M_{S_{\eta_c}}} \left|c \bar{c} \right\rangle\;,\\
T_2&=&g_s^2  \bar{c}_i t_{i j}^a \gamma_{\mu} c_j A_a^{\mu} \bar{c}_m t_{m n}^b \gamma_{\nu} c_n A_b^{\nu}\ .
\end{eqnarray}
Here, ${\textbf{k}}_c$ and ${\textbf{k}}_{\bar{c}}$ represent the $3$-momenta of quarks $c$ and $\bar{c}$; $\psi_{n_{\eta_c} L_{\eta_c} M_{L_{\eta_c}}} \left(\mathbf{k}_c, \mathbf{k}_{\bar{c}}\right)$ is the spatial wave function with $n$, $L$, $S$, $J$ the principal quantum number, orbital angular momentum, total spin and the total angular momentum of $|\eta_c \rangle$, respectively; $\chi^{c \bar{c}}$ is the corresponding spin state and $\langle L_G M_{L_G} S_G M_{S_G} | J_G M_{J_G} \rangle$ is the Clebsch-Gordan coefficient; $g_s$ denotes the strong coupling constant; $c_i$, $A^\mu_a$ and $t^a$ represent respectively the quark fields, gluon fields and Gell-Mann matrices.

Inserting the completeness relation $\sum_{G}| G \rangle\langle G |=2E_G$ into Eq.(\ref{transtiion}), we get
\begin{eqnarray}\label{T}
\langle G_1G_2|T|\eta_c\rangle &=& \frac{1}{2 E_G} \sum_{G} \gamma_g \langle G_1G_2 |{G_{\rho\sigma}^c G^{c\rho\sigma}}| G \rangle  \langle G |T_2 | \eta_c\rangle \nonumber \\
&\equiv& \frac{1}{2 E_G} \sum_{G} \gamma_g \langle G_1G_2 | T_1 | G \rangle \langle G | T_2 | \eta_c\rangle {+\text{high order terms}}\ ,
\end{eqnarray}
where $ | G \rangle $ is the shorthand notation for gluons $g_1$ and $g_2$ emitted from $\eta_c$ and the phase space integration is implied, as given in Eq.(\ref{midstateG}). $T_1$ stands for the operator responsible for $G\to G_1 G_2 $  transition.

\begin{figure}[htbp]
\begin{center}
\includegraphics[scale=0.46]{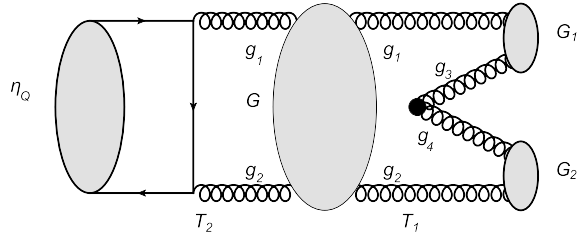}
\caption{The schematic Feynman Diagram of pseudoscalar quarkonium transition to glueball pair.}\label{T1T2}
\end{center}
\end{figure}

Noticing to evaluate the gluon-pair-vacuum interaction from the first principle, the QCD, is right now beyond our capability, we assume the interaction vertex shown in Fig. \ref{T1T2} can be modeled phenomenologically in such a way that the transition matrix $T_1$ is decomposed as:
\begin{eqnarray}
{T_1} = I_1 \otimes I_2 \otimes {T}_{vac}\ ,
\end{eqnarray}
where $T_{vac}$ signifies the vacuum-gluon pair transition operator, $I_i$ are identity matrices indicating the quasi-free propagations of $g_1$ and $g_2$. The gluons $g_3$ and $g_4$ are created in vacuum, and their spin states $ |m_{s_3},m_{s_4} \rangle $ having two different combinations. Please note that the gluons in transition matrix $T_1$ turn out to be massive, after experiencing some nonperturbative evolutions.

The total spin state of the gluon pair produced in vacuum, $ |S,M_S \rangle $, possessing the vacuum quantum number, being a singlet, can be formulated as
\begin{eqnarray}
\chi_{{0,0}}^{34} = \frac{1}{\sqrt{2}}\bigg(|1,-1 \rangle_{m_{s_3}m_{s_4}}+|-1,1 \rangle_{m_{s_3}m_{s_4}}\bigg)\ .
\end{eqnarray}
The ${T}_{vac}$ can then be expressed as
\begin{eqnarray}
{T}_{vac} &=& \gamma_g \int\!{\rm d^3}{\textbf{k}}_3\; {\rm d^3}{\textbf{k}}_4
\delta^3({\textbf{k}}_3+{\textbf{k}}_4) {\cal Y}_{00}
\left(\frac{{\textbf{k}}_3-{\textbf{k}_4}}{2} \right) \chi^{3 4}_{0, 0}\; \delta_{cd} a^\dagger_{3c}({\textbf{k}}_3) \;
a^\dagger_{4d}({\textbf{k}}_4)  \ . \label{tvac}
\end{eqnarray}
Here, ${\textbf{k}}_3$ and ${\textbf{k}}_4$ represent the $3$-momenta of gluons $g_3$ and $g_4$ respectively, $a^\dagger_{3c}$ and $a^\dagger_{4d}$ are creation operators of gluons with color indices, and $\mathcal{Y}_{\ell m}(\mathbf{k})\equiv |\mathbf{k}|^{\ell}Y_{\ell m}(\theta_{k},\phi_{k})$ is the $\ell$th solid harmonic polynomial that gives out the momentum-space distribution of the produced gluon pairs.

The state $| G \rangle$ should possess the quantum numbers of $| \eta_c\rangle$, i.e., $J^{PC}_G=0^{-+}$, which as discussed in Refs. \cite{Feldmann:1998vh,Feldmann:1998sh,Kochelev:2005tu,Tsai:2011dp} may have mixing with $\eta_c$, can thus parameterized as
\begin{eqnarray}\label{midstateG}
|G \rangle &=& \sqrt{2 E_G} \int \rm d^3 \mathbf{k}_1\rm d^3 \mathbf{k}_2 \delta^3\left(\textbf{K}_G-\mathbf{k}_1-{\mathbf{k}}_2\right)\nonumber\\
&\times & \sum_{M_{L_G},M_{S_G}} \left\langle L_G M_{L_G} S_G M_{S_G} | J_G M_{J_G} \right\rangle \psi_{n_G L_G M_{L_G}}\left(\mathbf{k}_1,\mathbf{k}_2\right) \chi^{1 2}_{S_G M_{S_G}}\delta_{ab} \left|g_1^a g_2^b \right\rangle\ ,
\end{eqnarray}
where ${\textbf{k}}_1$ and ${\textbf{k}}_2$ represent $3$-momenta of gluons $g_1$ and $g_2$, $\psi_{n_G L_G M_{L_G}} \left(\mathbf{k}_1, \mathbf{k}_2\right)$ is the spatial wave function with $n$, $L$, $S$, $J$ the principal quantum number, orbital angular momentum, total spin and the total angular momentum of $|G \rangle$, respectively. $\chi^{1 2}$ is the corresponding spin state, later
on expressed as $|S_G M_{S_G} \rangle$ for the sake of calculation transparency. $\langle L_G M_{L_G} S_G M_{S_G} | J_G M_{J_G} \rangle$ is the Clebsch-Gordan coefficient and reads $\langle 1m;1-m | 00 \rangle$ for $|G \rangle$ state. The normalization conditions write
\begin{eqnarray}\label{normalization1}
\langle G(\textbf{K}_G)|G(\textbf{K}'_G) \rangle &=& 2E_G\delta^3(\textbf{K}_G-\textbf{K}'_G)\ ,
\end{eqnarray}
\begin{eqnarray}\label{normalization2}
\langle g_i^a(\mathbf{k}_i)|g_j^b(\mathbf{k}_j)\rangle&=&\delta_{ij}\delta^{ab}\delta^3(\mathbf{k}_i-\mathbf{k}_j)\ ,
\end{eqnarray}
\begin{eqnarray}\label{normalization3}
\int \rm d^3 \mathbf{k}_1 \rm d^3 \mathbf{k}_2
\delta^3(\textbf{K}_G-\mathbf{k}_1-\mathbf{k}_2)
\psi_G(\mathbf{k}_1,\mathbf{k}_2)\psi_{G'}(\mathbf{k}_1,\mathbf{k}_2)=\delta_{G'G}\ ,~
\end{eqnarray}
with ${\textbf{K}_G}$ and ${\textbf{K}'_G}$ the corresponding $3$-momenta. Similarly we may have expressions for $G_1$ and $G_2$ states.

Equipped with the gluon-to-glueball transition operator $T_1$ and expressions for initial and final states, we are now capable of evaluating the transition matrix element
\begin{eqnarray}\label{T-matrix}
\langle G_1 G_2|T_1|G\rangle &=& \gamma_g\;\;\sqrt{8 E_G E_{G_1} E_{G_2}}\!\!\!\!\!\!\!\!\!\!\!
\sum_{\renewcommand{\arraystretch}{.5}\begin{array}[t]{l}
\scriptstyle (M_{L_G}, M_{S_G}), (\scriptstyle M_{L_{G_1}}, M_{S_{G_1}}), (\scriptstyle M_{L_{G_2}}, M_{S_{G_2}})
\end{array}}\renewcommand{\arraystretch}{1}\!\!\!\!\!\!\!\! \nonumber\\
&\times & \langle L_G M_{L_G} S_G M_{S_G} | J_G M_{J_G} \rangle\langle L_{G_1} M_{L_{G_1}} S_{G_1} M_{S_{G_1}} | J_{G_1} M_{J_{G_1}} \rangle\nonumber\\
&\times & \langle L_{G_2} M_{L_{G_2}} S_{G_2} M_{S_{G_2}} | J_{G_2} M_{J_{G_2}} \rangle \langle \chi^{1 3}_{S_{G_1} M_{S_{G_1}}}\chi^{2 4}_{S_{G_2} M_{S_{G_2}}}  | \chi^{1 2}_{S_G M_{S_G}} \chi^{3 4}_{0 0} \rangle\nonumber\\
&\times &
I_{M_{L_G},M_{L_{G_1}},M_{L_{G_2}}}({\textbf{K}})(\delta_{ab}\delta_{cd}\delta_{ac}\delta_{bd})_{color-octet} \; .
\end{eqnarray}
Here the momentum space integral $I_{M_{L_G},M_{L_{G_1}},M_{L_{G_2}}}({\textbf{K}})$ writes
\begin{eqnarray}
I_{M_{L_G},M_{L_{G_1}},M_{L_{G_2}}}({\textbf{K}}) &=& \int\!\rm
d^3 \mathbf{k}_1\rm d^3 \mathbf{k}_2\rm d^3 \mathbf{k}_3\rm
d^3 \mathbf{k}_4\,\delta^3(\mathbf{k}_1+\mathbf{k}_2 - \textbf{K}_G) \delta^3(\mathbf{k}_3+\mathbf{k}_4)\delta^3
(\textbf{K}_{G_1}-\mathbf{k}_1-\mathbf{k}_3)\nonumber\\
&\times &\delta^3(\textbf{K}_{G_2}-\mathbf{k}_2-\mathbf{k}_4) \psi^*_{n_{G_1} L_{G_1} M_{L_{G_1}}}(\mathbf{k}_1,\mathbf{k}_3)
\psi^*_{n_{G_2} L_{G_2} M_{L_{G_2}}}(\mathbf{k}_2\ ,\mathbf{k}_4) \nonumber\\
&\times &\psi_{n_G L_G M_{L_G}}(\mathbf{k}_1,\mathbf{k}_2)
\mathcal{Y}_{00}\Big(\frac{\mathbf{k}_3-\mathbf{k}_4}{2}\Big)\ .
\label{integral}
\end{eqnarray}

For simplicity, it is reasonable to conjecture the glueball and $|G\rangle$ state wave functions to be in harmonic oscillator (HO) form
\begin{eqnarray}
\psi_{nLM}(\mathbf{k})=\mathcal{N}_{nL}\exp\left(-\frac{R^2\mathbf{k}^2}{2}\right) \mathcal{Y}_{LM} (\mathbf{k})\,\mathcal{P}(\mathbf{k}^2)\ ,
\end{eqnarray}
where $\mathbf{k}$ is the relative momentum between two gluons inside the states, $\mathcal{N}_{nL}$ is the normalization coefficient and $\mathcal{P}(\mathbf{k}^2)$ is a polynomial of $\mathbf{k}^2$ \cite{Luo:2009}.
$\langle\chi^{1 3}_{S_{G_1} M_{S_{G_1}}}\chi^{2 4}_{S_{G_2} M_{S_{G_2}}}|\chi^{1 2}_{S_G M_{S_G}} \chi^{3 4}_{00} \rangle$ which denotes the spin coupling can be expressed in Wigner's $9j$ symbol \cite{yaouanc-book}
\begin{eqnarray}\label{angular}
\langle\chi^{1 3}_{S_{G_1} M_{S_{G_1}}}\chi^{2 4}_{S_{G_2} M_{S_{G_2}}}&|&\chi^{1 2}_{S_G M_{S_G}} \chi^{3 4}_{00} \rangle=(-1)^{S_{G_2}+1}\Big{[}(2S_{G_1}+1)(2S_{G_2}+1)(2S_G+1)\Big{]}^{1/2}\nonumber\\
&\times&\sum_{S,M_s}\langle
S_{G_1}M_{S_{G_1}};S_{G_2}M_{S_{G_2}}|SM_s \rangle \langle
SM_s|S_GM_{S_G};00 \rangle \nonumber\\
&\times&  \left\{ \begin{array}{ccc}
s_1 & s_3 & S_{G_1} \\
s_2 & s_4 & S_{G_2}\\
S_G & 0   & S
\end{array}
\right \} \;.
\end{eqnarray}
Here, $s_i$ is the spin of gluon $g_i$ with $i=1, 2, 3, 4, $ and $\sum_{S,M_s}|SM_s \rangle \langle SM_s|$ is the completeness relation.

The helicity amplitude $\mathcal{M}^{M_{J_G}M_{J_{G_1}}M_{J_{G_2}}}$ may be read off from
\begin{eqnarray}\label{M1}
\langle G_1 G_2|T_1|G\rangle=\delta^3(\mathbf{K}_{G_1}+\mathbf{K}_{G_2}-
\mathbf{K}_G)\mathcal{M}_1^{M_{J_G}M_{J_{G_1}}M_{J_{G_2}}}\ ,
\end{eqnarray}
and then the $\eta_c\to G_1G_2$ decay width can be readily obtained \cite{Luo:2009}:
\begin{eqnarray}\label{width}
\Gamma = \pi^2 \frac{{|\textbf{K}|}}{M_{\eta_c}^2}\sum_{JL}\Big|\mathcal{M}^{J L}\Big|^2\ .
\end{eqnarray}
Here, $\mathcal{M}^{J L}=\frac{\mathcal{M}_1^{J L} \mathcal{M}_2}{2 E_G}$,
$\mathcal{M}_2$ is the amplitude of process $\eta_c \to g g$, and
$\mathcal{M}_1^{J L}$ is the partial wave amplitude, obtainable from the helicity amplitude $\mathcal{M}_1^{M_{J_G}M_{J_{G_1}}M_{J_{G_2}}}$ via Jacob-Wick formula \cite{Jacob:1959}, i.e.
\begin{eqnarray}
{\mathcal{M}}_1^{J L} &=& \frac{\sqrt{2 L+1}}{2 J_G
+1} \!\! \sum_{M_{G_1},M_{G_2}} \langle L 0 J M_{J_G}|J_G
M_{J_G}\rangle \nonumber\\ \quad\quad\quad&\times&\langle J_{G_1}
M_{J_{G_1}} J_{G_2} M_{J_{G_2}} | J M_{J_G} \rangle \mathcal{M}_1^{M_{J_G}
M_{J_{G_1}} M_{J_{G_2}}}\label{coeff1}
\end{eqnarray}
with $J=J_{G_1}+J_{G_2}$ and $L=J_{G}-J$.

\section{Glueball pair production in pseudoscalar quarkonium decay}\label{decay}

In this section, we estimate the scalar and the pseudoscalar glueballs production
in $\eta_c$ and $\eta_b$ decays via the $0^{++}$ model, by taking scalars $f_0(1710)$ and $f_0(1500)$, and pseudoscalar $\eta(1405)$ as the corresponding candidates, namely $G_1$ and $G_2$ respectively. The quantum numbers of the states involve in these processes are presented in Table~\ref{tab1}, where $|G\rangle$ and $|\eta_Q\rangle$ have the same quantum numbers.

\begin{table}[htbp]
\caption{Quantum numbers of $\eta_Q$, $G_1$ and $G_2$. The values of $M_0$ and $M_2$ can be $-1$, $0$ and $1$.}
\begin{center}\label{tab1}
\begin{tabular}{|c|c|c|c|c|c|c|c}\hline\hline
                 & $J^{PC}$   & $L$ & $M_L$     & $S$  & $M_S$           \\ \hline
 $\eta_Q    $    & $0^{-+}$   & $1$ & $M_0$     & $1$  & $-M_0$          \\ \hline
 $G_1$    & $0^{++}$   & $0$ & $0$       & $0$  & $0$             \\ \hline
 $G_2$    & $0^{-+}$   & $1$ & $M_2$     & $1$  & $-M_2$          \\ \hline
 \hline
\end{tabular}
\end{center}
\end{table}

\subsection{The evaluation of $T_1$}

In Eq.(\ref{T-matrix}), the color contraction yields number $8$, and for scalar glueballs,
the spin and orbital angular momentum coupling leads the C-G coefficient to be
$\langle 00;00| 00\rangle=1$. Therefore, Eq.(\ref{T-matrix}) in this situation turns to
\begin{eqnarray}\label{T-matrix1}
\langle G_1 G_2|T_1|G\rangle &=& \sum_{M_G,M_{G_2}}8 \gamma_g  \sqrt{8 E_G E_{G_1} E_{G_2}}
 \langle 1M_0;1-M_0 | 00 \rangle
\langle 1M_2;1-M_2 | 00 \rangle \nonumber\\
&\times &  \langle\chi^{1 3}_{0 0}\chi^{2 4}_{1 -M_2}  |
\chi^{1 2}_{1 -M_0} \chi^{3 4}_{0 0} \rangle I_{M_0,0,M_2}({\textbf{K}})\;.
\end{eqnarray}
The spin coupling $\langle\chi^{1 3}_{0 0}\chi^{2 4}_{1 -M_2} | \chi^{1 2}_{1 -M_0}
\chi^{3 4}_{0 0} \rangle$ is characterized by the Wigner's $9j$ symbol,
a representation of $4$-particle spin coupling, which can be expanded as series of
$2$-particle spin couplings represented by Wigner's $3j$ symbols \cite{yaouanc-book}, shown in Appendix \ref{appendixa}.

By substituting the spin couplings given in Appendix \ref{appendixa} into Eq.(\ref{T-matrix1}), we can then reduce the $T_1$ matrix element,
\begin{eqnarray}\label{T-matrix2}
\langle G_1 G_2|T_1|G\rangle &=& -\frac{1}{6}\gamma_g \sqrt{8 E_G E_{G_1} E_{G_2}}\nonumber\\
&\times&\bigg(|\langle11,1-1|00\rangle|^2 I_{1,0,1}(\textbf{K})
+|\langle10,10|00\rangle|^2 I_{0,0,0}
+|\langle1-1,11|00\rangle|^2 I_{-1,0,-1}(\textbf{K})\bigg)\nonumber\\
&=&-\frac{\gamma_g}{18}\sqrt{8 E_G E_{G_1} E_{G_2}}
\bigg(I_{1,0,1}(\textbf{K})+I_{0,0,0}(\textbf{K})+I_{-1,0,-1}(\textbf{K})\bigg)\ .
\end{eqnarray}
With a lengthy calculation, some details are given in Appendix \ref{appendixb}, the momentum space integrals are obtained, of which
$I_{1,0,1}=I_{-1,0,-1}=0$, and $I_{0,0,0}$ is given in Eq. (\ref{integral0000}). Giving  $\delta^3(\textbf{K}_G-\textbf{K}_{G_1}-\textbf{K}_{G_2})I
\equiv I_{0,0,0}$ and considering Eqs. (\ref{M1}), (\ref{T-matrix2}) and (\ref{integral0000}), we have
\begin{eqnarray}\label{finalinte}
\langle G_1 G_2| {T_1}|G\rangle&=&\delta^3(\textbf{K}_G-\textbf{K}_{G_1}-
\textbf{K}_{G_2})\mathcal{M}_1^{M_{J_G}M_{J_{G_1}}M_{J_{G_2}}}\nonumber\\
&=&-\frac{\gamma_g}{18}\sqrt{8 E_G E_{G_1} E_{G_2}}I_{0,0,0}\nonumber\\
&=&-\frac{\gamma_g}{18}\sqrt{8 E_G E_{G_1} E_{G_2}}\delta^3(\textbf{K}_G-\textbf{K}_{G_1}-\textbf{K}_{G_2})\ I\ ,
\end{eqnarray}
from which $\mathcal{M}_1^{M_{J_G}M_{J_{G_1}}M_{J_{G_2}}}=\mathcal{M}_1^{000}$ can be extracted out, i.e.
\begin{eqnarray}
\mathcal{M}_1^{000}=-\frac{\gamma_g}{18}\ I\ \sqrt{8 E_G E_{G_1} E_{G_2}}\ \ .
\end{eqnarray}

The probable radius $R$ of the HO wave function is estimated relation $R=1/\alpha$, with
$\alpha = \sqrt{\mu \omega /\hbar}$. Here, $\mu$ denotes the reduced mass, $\omega$ is
the angular frequency of harmonic oscillator satisfying $M = (2n+L+3/2)\hbar\omega$,
with $M$ being the glueball mass, $n$ the radial quantum number, and $L$ the orbital
angular momentum. As discussed in Refs.\cite{Bernard:1982,Cornwall:1982}, the effective mass of the constituent gluon is about $0.6$ GeV, which means $\mu \sim 0.3$ GeV for glueballs. In the calculation, the inputs we adopt are: $M_{\eta_c}=2.98$ GeV, $M_{\eta_b}=9.40$ GeV, $M_{f_0(1500)}=1.50$ GeV, $M_{f_0(1710)}=1.71$ GeV and $M_{\eta(1405)}=1.41$ GeV \cite{pdg}. Therefore, the corresponding radii $R_{\eta_c}=2.24\;\text{GeV}^{-1}$, $R_{\eta_b}=1.26\;\text{GeV}^{-1}$, $R_{f_0(1500)}=2.79\;\text{GeV}^{-1}$, $R_{f_0(1710)}=2.61\;\text{GeV}^{-1}$ and $R_{\eta(1405)}=3.26\;\text{GeV}^{-1}$.

With above discussion and inputs, we can readily get $I$ and $\mathcal{M}_1^{000}$. Please note that, when
\begin{eqnarray}
\langle L 0 J M_{J_G}|J_G M_{J_G}\rangle = \langle L 0 J 0|0 0\rangle=
\langle 0 0 0 0|0 0\rangle=1\; ,\\
\langle J_{G_1}M_{J_{G_1}}J_{G_2} M_{J_{G_2}} | J M_{J_G}\rangle=\langle0 0 0 0|00 \rangle=1 \; ,
\end{eqnarray}
$\mathcal{M}_1^{0 0}$ can be obtained according to Eq. (\ref{coeff1}), as shown in Table~\ref{tab2}.

\begin{table}[htbp]
\caption{ The $I$ and $\mathcal{M}_1^{0 0}$ in difference processes.}
\begin{center}\label{tab2}
\begin{tabular}{|c|c|c|c|c|c|c|c|}\hline\hline
                                                       & $I$ $(\text{GeV})^{-3/2}$  & $\mathcal{M}_1^{0 0}$   \\ \hline
 $\eta_c\to f_0(1500)\eta(1405)$                         & $0.409$                      &$-0.166\gamma_g$                \\ \hline
 $\eta_b\to f_0(1500)\eta(1405)$                        & $-0.398$                      &$0.901\gamma_g$                \\ \hline
 $\eta_b\to f_0(1710)\eta(1405)$                        & $-0.396$                      &$0.897\gamma_g$                \\ \hline
 \hline
\end{tabular}
\end{center}
\end{table}


\subsection{The evaluation of $T_2$}
\begin{figure}[htb]
\begin{center}
\includegraphics[scale=0.46]{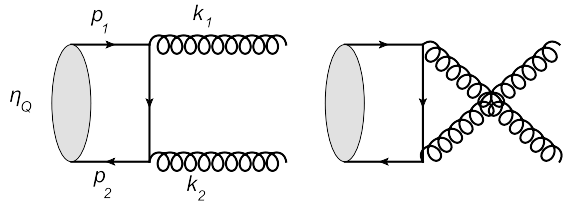}
\caption{The Feynman diagrams of $\eta_Q \to gg$ decay process.}\label{etaggfig}
\end{center}
\end{figure}
The calculation of the process $\eta_Q \to gg$ is quite straightforward. At the leading order
of perturbative QCD, there are only two Feynman diagrams, shown in Fig. \ref{etaggfig}. Their decay amplitudes write directly:
\begin{eqnarray}
i \mathcal{A}^{\mu\nu,ab}_1 \epsilon^{\ast}_{\mu}(k_1)\epsilon^{\ast}_{\nu}(k_2)&=&(ig_s)^2
\bar{v}(p_2)\gamma^\nu t^b \frac{i}{{p\!\!\!/}_1-{k\!\!\!/}_1-m_Q} \gamma^\mu t^a u(p_1)\epsilon^{\ast}_{\mu}(k_1)\epsilon^{\ast}_{\nu}(k_2)\; ,\\
i \mathcal{A}^{\mu\nu,ab}_2 \epsilon^{\ast}_{\mu}(k_1)\epsilon^{\ast}_{\nu}(k_2)&=&(ig_s)^2 \bar{v}(p_2)\gamma^\mu t^a \frac{i}{{p\!\!\!/}_1-{k\!\!\!/}_2-m_Q} \gamma^\nu t^b u(p_1)\epsilon^{\ast}_{\mu}(k_1)\epsilon^{\ast}_{\nu}(k_2)\; ,
\label{etagg}
\end{eqnarray}
where $u$ and $\bar{v}$ stand for heavy quark spinors, $\epsilon_\mu$ denotes
gluon polarization, and $g_s$ is the strong coupling constant. For quark pair
to form a pseudoscalar quarkonium, in normal routine one can realize it by performing
the following projection \cite{Qiao:2007}:
\begin{eqnarray}
u(p)\bar{v}(-p)\to{i\gamma_5 R_{\eta_Q}(0)\over 2
\sqrt{2\pi\times m_Q}} \,(p\!\!\!/+ m_Q)\,\otimes \left( {{\bf
1}_c\over \sqrt{N_c}}\right)\  \label{Etab:projector}\ .
\end{eqnarray}
Here, $m_Q$ is the heavy quark mass, $R_{\eta_Q}(0)$ denotes radial wavefunction at the origin, and in $\eta_Q$ center-of-mass system $p_1=p_2\equiv p$. The $\eta_Q \to gg$ matrix element squared may be obtained through a straightforward calculation, i.e.
\begin{eqnarray}
|\mathcal{M}_2|^2=\frac{4 g_s^4 |R(0)_{\eta_Q}|^2}{3 \pi m_Q}\;.
\end{eqnarray}

\subsection{The estimation of $\gamma_g$}

\begin{figure}[htbp]
\begin{center}
\includegraphics[scale=0.46]{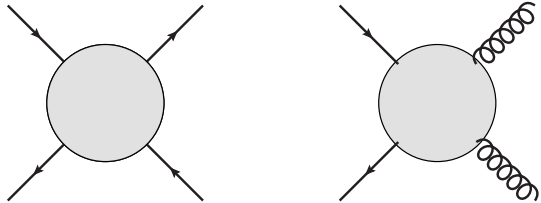}
\caption{The coupling of $q\bar{q}q\bar{q}$ and $q\bar{q}gg$.}\label{coupling}
\end{center}
\end{figure}

We estimate the strength of gluon-pair-vacuum coupling in analogous to the $^3P_0$ model, where the strength of quark pair creation in vacuum is represented by $\gamma_q$ with energy dimension \cite{Segovia:2012}. To avoid constructing a new model to mimic the non-perturbative process of the gluon pair production in the vacuum, we simply infer the $\gamma_g$ by comparing the relative strength of processes $q\bar{q}\to gg$ and $q\bar{q} \to q\bar{q}$, as shown in Fig.\ref{coupling}. The $\gamma_g^2/\gamma_q^2$ is assumed to be at the same order of magnitude as the relative interaction rate of these two processes.

It is well known that at the tree level
\begin{eqnarray}
|\bar{M}({q\bar{q}\to q\bar{q}})|^2 &=&  \frac{4 g_s^4}{9} \left(\frac{s^2+u^2}{t^2}+\frac{t^2+u^2}{s^2}-\frac{2 u^2}{3 s t}\right)\,,\\
|\bar{M}({q\bar{q}\to gg})|^2 & = & \frac{32 g_s^4}{27} \left(-\frac{9 \left(t^2+u^2\right)}{4 s^2}+\frac{t}{u}+\frac{u}{t}\right)\,,
\end{eqnarray}
and considering of the relationship between Mandelstam variables, we get
\begin{eqnarray}
{\gamma_g^2}/{\gamma_q^2} \approx \frac{\sigma({q\bar{q}\to gg})}{\sigma ({q\bar{q}\to q\bar{q}})}\approx (1.10\pm0.37)\times 10^{-2}\ ,
\end{eqnarray}
where the interaction energy is set to be $\mu_{\eta_c}$, the reduced mass of the quark$-$antiquark in the decaying meson.
In $^3P_0$ model $\gamma_q (\mu_{\eta_c}) = 0.299\times 2 m_q \sqrt{96\pi}$ \cite{Segovia:2012} with $m_q = 220\ \mathrm{MeV}$ \cite{Sun:2014} being the light quark constituent mass, and hence we get $\gamma_g^2(\mu_{\eta_c}) \approx (5.74\pm1.93)\times10^{-2}\ \mathrm{GeV}^{2}$. Similarly, we obtain $\gamma_g^2(\mu_{\eta_b}) \approx (2.57\pm0.86)\times10^{-3}\ \mathrm{GeV}^{2}$.

\subsection{Glueball production rate in $0^{++}$ model}

Eq.(\ref{T}) tells $\mathcal{M}^{J L}=\frac{\mathcal{M}_1^{J L}
\mathcal{M}_2}{2 E_G}$, where $\mathcal{M}^{J L}$ has only one nonzero matrix element, the $\mathcal{M}_1^{00}=-\frac{\gamma_g}{18}I\sqrt{8 E_G E_{G_1} E_{G_2}}$, and $|\mathcal{M}_2|^2=\frac{4 g_s^4 |R(0)_{\eta_Q}|^2}{3 \pi m_Q}$. Substituting them into Eq.(\ref{width}), we can then get the $\eta_c \to f_0(1500) \eta(1405)$ decay width
\begin{eqnarray}
\Gamma&=&\pi^2 \frac{{|\textbf{K}|}}{M_{\eta_c}^2}\sum_{JL}\Big|\mathcal{M}^{J L}\Big|^2=\pi^2\frac{|\mathbf{K}|}{4 M_{\eta_c}^4}|\mathcal{M}_1^{0 0}|^2 |\mathcal{M}_2|^2 \nonumber\\
&=&\frac{{2} \pi^2 g_s^4 |R(0)_{\eta_c}|^2 \gamma_g^2 |\mathbf{K}| E_G E_{G_1} E_{G_2}
I^2}{{3^5} \pi m_c {M_{\eta_c}^4 }}\nonumber\\
&=&{27.41^{+11.02}_{-10.12}\ \mathrm{keV}}\;.
\end{eqnarray}
In above calculation, we adopt the charm quark mass $m_c = (1.27 \pm 0.03)\ \mathrm{GeV}$ \cite{pdg},
strong coupling constant $\alpha_s(\eta_c)=0.25$, and the $\eta_c$ radial wave function at the origin squared $|R(0)_{\eta_c}|^2 = 0.527\pm0.013$ ${\rm GeV}^{3}$ \cite{Qiao:2007}. The branching fraction of $\eta_c \to f_0(1500)\eta(1405)$ process is then
\begin{eqnarray}
Br_{\eta_c \to f_0(1500)\eta(1405)}=\frac{\Gamma_{\eta_c \to f_0(1500)
\eta(1405)}}{\Gamma_{total}}={8.62^{+3.77}_{-3.32}\times10^{-4}}\ .~~~
\end{eqnarray}

In analogous to the $\eta_c$ decay, the $\eta_b$ exclusive decay to glueball pairs can be evaluated by $0^{++}$ model as well. We notice that $f_0(1710)$ is glue rich \cite{Janowski:2014ppa,Brunner:2015oqa,Gui:2012gx}, and evaluate as well the process $\eta_b\to f_0(1710)\eta(1405)$. With the same procedure performed for $\eta_c$, we have
\begin{eqnarray}
\Gamma_{\eta_b\to f_0(1500)\eta(1405)}= {7.57^{+2.68}_{-2.60}} \ \mathrm{keV}\ ,\; Br_{\eta_b\to f_0(1500)\eta(1405)} = {7.57^{+9.50}_{-4.26}\times 10^{-4}}\ \; .
\end{eqnarray}
\begin{eqnarray}
{\Gamma_{\eta_b\to f_0(1710)\eta(1405)} = {7.34^{+2.60}_{-2.53}} \ \mathrm{keV}\ ,\; Br_{\eta_b\to f_0(1710)\eta(1405)} = { 7.35^{+9.23}_{-4.14}\times 10^{-4}}}\ \; .
\end{eqnarray}
Here in the calculation, the bottom quark mass $m_b = (4.18 \pm 0.03)\ \mathrm{GeV}$ \cite{pdg}, the strong coupling constant $\alpha_s(\eta_b)=0.18$, and the $\eta_b$ radial wave function at the origin squared $|R(0)_{\eta_b}|^2 = 4.89\pm0.07$ ${\rm GeV}^{3}$ \cite{Qiao:2007} are adopted. It is worthwhile to mention that though there are mixings among $f_0(1370)$, $f_0(1500)$ and $f_0(1710)$ states \cite{Janowski:2014ppa}, they do not influence much on our calculation results.

Moreover, from the Lattice QCD calculations
\cite{Lee:1999,Bali:1993,Morningstar:1997,Chen:2006,Morningstar:1999},
we know that there might be scalar and pseudoscalar glueball candidates with masses
$1.75$ GeV and $2.39$ GeV, respectively. In this situation we can readily get
\begin{eqnarray}
\Gamma_{\eta_b\to G^{0^{++}}G^{0^{-+}}} = {4.56^{+1.61}_{-1.57}} \ \mathrm{keV}\ ,\; Br_{\eta_b\to G^{0^{++}}G^{0^{-+}}} = {4.56^{+5.72}_{-2.56}\times 10^{-4}}\ \; .
\end{eqnarray}

\section{Summary}

In this work, we analyze the processes of glueball pair exclusive production in quarkonium decays by introducing a $0^{++}$ model, which is employed to mimic phenomenologically the gluon-pair-vacuum interaction vertices and is applicable to studies of glueball and hybrid state production. It is assumed that gluon pair is created homogeneously in space with equal probability. By virtual of the $^3P_0$ model, we formulate an explicit vacuum gluon-pair transition matrix and estimate the strength of the gluon-pair creation. We then apply it to the calculation of $\eta_c$ to $f_0(1500)$ $\eta(1405)$ decay process, where $f_0(1500)$ and $\eta(1405)$ are supposed to be scalar and pseudoscalar glueball candidates respectively, and find the decay width and branching ratio are ${27.41}$ keV and ${8.62\times 10^{-3}}$ respectively.

In light of the $\eta_c$ decay, we evaluate also the $\eta_b\to f_0(1500)\eta(1405)$ and $\eta_b\to f_0(1710)\eta(1405)$ processes, and find the decay widths and branching ratios are $7.57$ keV and $7.57\times 10^{-4}$; $7.34$ keV and $7.35\times 10^{-2}$, respectively. Suppose there exit heavier scalar and pseudoscalar glueballs with masses $1.75$ GeV and $2.39$ GeV as per the Lattice QCD calculation, we find the corresponding decay width and branching ratio are ${4.56}$ keV and ${4.56\times 10^{-4}}$.
Our results in this work indicate that the glueball pair production in pseudoscalar quarkonium decays is marginally accessible in the presently running experiments BES III, BELLE II, and LHCb.

It should be mentioned that the hadronic two-body decay modes of the scalar-isoscalar $f_0(1370)$, $f_0(1500)$ and $f_0(1710)$ were investigated in Ref.\cite{StrohmeierPresicek:1999yv}, where the leading order process $G_0\to G_0G_0$ was also proposed, but neglected in practical caulculation. We believe that, in future study, the combination of $0^{++}$ model with the analysis in Ref.\cite{StrohmeierPresicek:1999yv} would no doubt informs us more about the properties of glueballs and isoscalar mesons.

Last, we acknowledge that the estimation for gluon-pair-vacuum coupling here is quite premature and hence the estimation for pseudoscalar quarkonium exlusive decay to glueballs is far from accurate. However, qualitatively the physical picture is naively sound. To make the $0^{++}$ mechanism trustworthy in phenomenology study, or in other words to ascertain the coupling strength, a first step experimental measurement would be on $\eta_c \to \eta'(958) + f_0(1500)$ process, since we know the $\eta'(958)$ is also a glue rich object. With the increase of experimental measurement on glueball production and decay, the model will be refined, and hence its predictability, which no doubt requires a lot of tedious works. However, due to the importance of the glueball physics, we believe it deserves to explore.


\section*{Acknowledgements}
Authors are grateful to anonymous reviewers' comments and suggestions, which are important for the completeness and improvement of the paper. This work was supported in part by the Strategic Priority Research Program of the Chinese Academy of Sciences, Grant No.XDB23030100; and by the National Natural Science Foundation of China(NSFC) under the Grants 11975236 and 11635009.

\appendix

\section{Wigner's symbols}\label{appendixa}
In Eq.(\ref{T-matrix1}), the Wigner's $3j$ and $9j$ symbols are
\begin{eqnarray}
\left \{\begin{array}{ccc}
j_1 & j_2 & j \\
m_1 & m_2 & m
\end{array}
\right \}=\frac{(-1)^{j_1-j_2-m}}{\sqrt{2j+1}}\langle j_1 j_2 m_1 m_2|j,-m\rangle
\end{eqnarray}
and
\begin{eqnarray}
\left \{\begin{array}{ccc}
j_1    & j_2    & j_{12} \\
j_3    & j_4    & j_{34} \\
j_{13} & j_{24} & j
\end{array}
\right \}&=&\sum\limits_{m}
\left \{\begin{array}{ccc}
j_1 & j_2 & j_{12} \\
m_1 & m_2 & m_{12}
\end{array}
\right \}
\left \{\begin{array}{ccc}
j_3 & j_4 & j_{34} \\
m_3 & m_4 & m_{34}
\end{array}
\right \}
 \left \{\begin{array}{ccc}
j_{13} & j_{24} & j \\
m_{13} & m_{24} & m
\end{array}
\right \}\nonumber\\
&\times&
\left \{\begin{array}{ccc}
j_1 & j_3 & j_{13} \\
m_1 & m_3 & m_{13}
\end{array}\right \}\left \{\begin{array}{ccc}
j_2 & j_4 & j_{24} \\
m_2 & m_4 & m_{24}
\end{array}\right \}
\left \{\begin{array}{ccc}
j_{12} & j_{34} & j \\
m_{12} & m_{34} & m
\end{array}\right \}\ ,~~~~~~
\end{eqnarray}
respectively. Applying them to Eq.(\ref{angular}) reduces the spin coupling term to
\begin{eqnarray}\label{angular1}
\langle \chi^{1 3}_{0 0}\chi^{2 4}_{1 -M_2}  | \chi^{1 2}_{1 -M_0} \chi^{3 4}_{0 0} \rangle=3\sum\limits_{S,M_S}\langle00;1-M_2|SM_S\rangle \langle SM_S|1-M_0;00\rangle\left \{\begin{array}{ccc}
1    & 1    & 0 \\
1    & 1    & 1 \\
1    & 0    & S
\end{array}
\right \}\;.
\end{eqnarray}
In above equation, evidently $\langle00;1-M_2|SM_S\rangle$ and $\langle SM_S|1-M_0;00 \rangle$
become nonzero only when $S=1$, which means $M_S$ can be any of $1$, $0$ or $-1$.
Thus all possible $|SM_S\rangle$ are $|1,-1\rangle$, $|1,0\rangle$ and $|1,1\rangle$.
On the other hand, $\langle00;1-M_2|SM_S\rangle$ and $\langle SM_S|1-M_0;00\rangle$ will be zero unless $M_0=M_2=-M_S$.

Given $M \equiv M_S$ the Wigner's $9j$ symbol can then be calculated as follows:
\begin{eqnarray}
\left \{\begin{array}{ccc}
1    & 1    & 0 \\
1    & 1    & 1 \\
1    & 0    & 1
\end{array}
\right \} &=&\sum\limits_{M}
\left \{\begin{array}{ccc}
1   & 1   & 0 \\
m_1 & m_3 & 0
\end{array}
\right \}
\left \{\begin{array}{ccc}
1   & 1   & 1 \\
m_2 & m_4 & -M
\end{array}
\right \}
\left \{\begin{array}{ccc}
1  & 0 & 1 \\
M  & 0 & -M
\end{array}
\right \}\nonumber\\
&\times&
 \left \{\begin{array}{ccc}
1   & 1   & 1 \\
m_1 & m_2 & -M
\end{array}\right \}
\left \{\begin{array}{ccc}
1   & 1   & 0 \\
m_3 & m_4 & 0
\end{array}\right \}
\left \{\begin{array}{ccc}
0 & 1  & 1 \\
0 & M  & -M
\end{array}\right \}\nonumber\\
&=&\frac{1}{9}\langle1m_1;1m_3|00\rangle\langle1m_2;1m_4|1 M\rangle\langle1M;00|1 M\rangle\nonumber\\
&\times&\langle1m_1;1m_2|1 M\rangle \langle1m_3;1m_4|00\rangle\langle00;1M|1 M\rangle\ .
\end{eqnarray}
Provided only transverse polarization exists, every term in above equation can be evaluated in normal C-G coefficient. That is,
\begin{eqnarray}
\langle1m_1;1m_3|00\rangle = \sqrt{\frac{1}{2}}(\delta_{m_1 1}\delta_{m_3, -1} - \delta_{m_1, -1}\delta_{m_3 1})\ ,~~
\end{eqnarray}
\begin{eqnarray}
\langle1M;00|1M\rangle = \frac{\sqrt{2}}{2}\ ,
\end{eqnarray}
\begin{eqnarray}
\langle00;1M|1M\rangle = \frac{\sqrt{2}}{2}\ ,
\end{eqnarray}
\begin{eqnarray}
\langle1m_3;1m_4|00\rangle = \sqrt{\frac{1}{2}}(\delta_{m_3 1}
\delta_{m_4, -1}-\delta_{m_3, -1}\delta_{m_4 1})\; ,~~~
\end{eqnarray}
\begin{eqnarray}
\langle1m_2;1m_4|1-1\rangle =0 \ ,
\end{eqnarray}
\begin{eqnarray}
\langle1m_1;1m_2|1-1\rangle =0 \ ,
\end{eqnarray}
\begin{eqnarray}
\langle1m_2;1m_4|10\rangle = \frac{\sqrt{2}}{2}(\delta_{m_2 1}\delta_{m_4, -1}-\delta_{m_2, -1}\delta_{m_4 1})\ ,
\end{eqnarray}
\begin{eqnarray}
\langle1m_1;1m_2|10\rangle = \frac{\sqrt{2}}{2}(\delta_{m_1 1}\delta_{m_2, -1}-\delta_{m_1, -1}\delta_{m_2 1})\ ,
\end{eqnarray}
\begin{eqnarray}
\langle1m_2;1m_4|11\rangle =0\ ,
\end{eqnarray}
\begin{eqnarray}
\langle1m_1;1m_2|11\rangle =0\ .
\end{eqnarray}

After inserting above ingredients into Eq.(\ref{angular1}), we get the only nonzero spin coupling as
\begin{eqnarray}
\langle \chi^{1 3}_{0 0}\chi^{2 4}_{1 0}  | \chi^{1 2}_{1 0} \chi^{3 4}_{0 0} \rangle &=& \frac{1}{48}(\delta_{m_1 1}\delta_{m_3, -1}-\delta_{m_1, -1}\delta_{m_3 1}) (\delta_{m_3 1}\delta_{m_4, -1}-\delta_{m_3, -1}\delta_{m_4 1})\nonumber\\
&\times& (\delta_{m_2 1}\delta_{m_4, -1}-\delta_{m_2, -1}\delta_{m_4 1}) (\delta_{m_1 1}\delta_{m_2, -1}-\delta_{m_1, -1}\delta_{m_2 1})\;,
\end{eqnarray}
which equals $-\frac{1}{48}$ for $m_1=-1$, $m_2=1$, $m_3=1$, $m_4=-1$ or $m_1=1$, $m_2=-1$, $m_3=-1$, $m_4=1$, and $0$ for all other cases.

\section{The momentum space integrals}\label{appendixb}
For non-trivial situation, that is $M_0=M_2=-M$, the momentum integral $I_{M_{L_G},M_{L_{G_1}},M_{L_{G_2}}}({\textbf{K}})$ in Eq.(\ref{T-matrix2}) reduces to
\begin{eqnarray}
I_{M,0,M}(\textbf{K}) &=& \int\!\rm
d^3 \mathbf{k}_1\rm d^3 \mathbf{k}_2\rm d^3 \mathbf{k}_3\rm
d^3 \mathbf{k}_4\,\delta^3(\mathbf{k}_1+\mathbf{k}_2)
 \delta^3(\mathbf{k}_3+\mathbf{k}_4)\delta^3(\textbf{K}_{G_1}-\mathbf{k}_1-\mathbf{k}_3)
 \delta^3(\textbf{K}_{G_2}-\mathbf{k}_2-\mathbf{k}_4)\nonumber\\
&\times &\psi^*_{n_1 0 0}(\mathbf{k}_1,\mathbf{k}_3) \psi^*_{n_2 1 M}(\mathbf{k}_2,\mathbf{k}_4) \psi_{n_0 1 M}(\mathbf{k}_1,\mathbf{k}_2)\mathcal{Y}_{00}\Big(\frac{\mathbf{k}_3-\mathbf{k}_4}{2}\Big)\ .
\label{integral1}
\end{eqnarray}
Provided the ground state dominance holds, namely the principal numbers $n_0$, $n_1$ and $n_2$ equal to $1$, the wave function $\psi$ then turns to
\begin{eqnarray}\label{sho1}
\psi_{100}(\mathbf{k}) = \frac{1}{\pi^{3/4}}R^{3/2}\exp\left(-\frac{R^2\mathbf{k}^2}{2}\right)\ ,
\end{eqnarray}
\begin{eqnarray}\label{sho2}
\psi_{11M}(\mathbf{k}) = i\frac{\sqrt{2}}{\pi^{3/4}}R^{5/2}k_M\exp\left(-\frac{R^2\mathbf{k}^2}{2}\right)\ ,
\end{eqnarray}
where $k_M$, $k_{\pm1}=\mp(k_x\pm{}ik_y)/\sqrt{2}$ and $k_{0}=k_z$ are the spherical components of vector
$\mathbf{k}$.

Integrating out those dummy variables, we can simplify Eq.(\ref{integral1}) like
\begin{eqnarray}
I_{M,0,M}(\textbf{K}) &=& \delta^3(\textbf{K}_G-\textbf{K}_{G_1}-
\textbf{K}_{G_2}) \int\!\rm d^3 \mathbf{k}_1
\psi^{1*}_{1 0 0}(\mathbf{k}_1,\textbf{K}-\mathbf{k}_1)
\nonumber\\
&\times &\psi^{2*}_{1 1 M}(-\mathbf{k}_1,-\textbf{K}+\mathbf{k}_1)\psi^G_{1 1 M}(\mathbf{k}_1,-\mathbf{k}_1)
\mathcal{Y}_{00}(\mathbf{k}_1)\;.
\end{eqnarray}
In addition, in the $\eta_Q$ center-of-mass system, which implies $\textbf{K}_G=
\textbf{K}_{\eta_Q}=0$ and $\textbf{K}_{G_1}=-\textbf{K}_{G_2}=\textbf{K}$,
the spatial wave functions given in (\ref{sho1}) and (\ref{sho2}) may be written as
\begin{eqnarray}
\psi^{1*}_{1 0 0} = \frac{R_1^{3/2}}{\pi^{3/4}}exp\big(-\frac{R_1^2(2\mathbf{k}_1-\textbf{K})^2}{8}\big)\ ,
\end{eqnarray}
\begin{eqnarray}
\psi^{2*}_{1 1 M} = -i\frac{R_2^{5/2}}{\sqrt{2}\pi^{3/4}}(2\mathbf{k}_1-\textbf{K})_M \ exp\big(-\frac{R_2^2(2\mathbf{k}_1-\textbf{K})^2}{8}\big)\ ,
\end{eqnarray}
\begin{eqnarray}
\psi^G_{1 1 M} = i\frac{\sqrt{2}R_0^{5/2}}{\pi^{3/4}}(\mathbf{k}_1)_M\ exp\big(-\frac{R_0^2(\mathbf{k}_1)^2}{2}\big)\ ,
\end{eqnarray}
with $\mathcal{Y}_{00} = \frac{1}{\sqrt{4\pi}}$. Here, $R_0$, $R_1$ and $R_2$ are the most probable radii of $\eta_c$, $f_0(1500)$ and $\eta(1405)$, respectively.
After performing the integration, one may find that the states
$M=1$ and $M=-1$ do not make any contribution, i.e. $I_{1,0,1}=I_{-1,0,-1}=0$, while
 \begin{eqnarray}\label{integral0000}
I_{0,0,0}&=&-\delta^3(\textbf{K}_G-\textbf{K}_{G_1}-\textbf{K}_{G_2})\frac{R_1^{3/2}R_2^{5/2}R_0^{5/2}}{6 \sqrt{2}\pi^{5/4}(R_0^2+R_1^2+R_2^2)^{9/2}}
exp\bigg(-\frac{\textbf{K}^2R_0^2(R_1^2+R_2^2)}{8(R_0^2+R_1^2+R_2^2)}\bigg)\nonumber\\
&\times&\bigg\{R_0^2 \left(R_1^2+R_2^2\right) \left[\textbf{K}^4 \left(R_1^2+R_2^2\right)^2-96\right]
+12 R_0^4 \left[\textbf{K}^2 \left(R_1^2+R_2^2\right)-4\right]\nonumber\\
&-&12 \left(R_1^2+R_2^2\right)^2 \left[\textbf{K}^2 \left(R_1^2+R_2^2\right)+4\right]\bigg\}\;.
\end{eqnarray}

\end{document}